%% file: main.tex
\documentclass[12pt]{iopart}

\input{header}

\begin{document}

\title{Palladium Zero-Mode Waveguides for Optical Single Molecule Detection with Nanopores}

\author{Nils Klughammer and Cees Dekker}

\address{Department of Bionanoscience, Kavli Institute of Nanoscience, Delft University of Technology, Van der	Maasweg 9, 2629 HZ, Delft, Netherlands}
\ead{c.dekker@tudelft.nl}
\vspace{10pt}

\begin{abstract}
Holes in metal films block any transmitting light if the wavelength is much larger than the hole diameter, establishing such nanopores as so-called Zero Mode Waveguides (ZMWs). Molecules on the other hand, can still passage through these holes. We use this  to detect individual fluorophore-labelled molecules as they travel through a ZMW and thereby traverse from the dark region to the illuminated side, upon which they emit fluorescent light. This is beneficial both for background suppression and to prevent premature bleaching. We use palladium as a novel metal-film material for ZMWs, which is advantageous compared to conventionally used metals. We demonstrate that it is possible to simultaneously detect translocations of individual free fluorophores of different colors. Labeled DNA and protein biomolecules can be detected as well at the single-molecule level with a high signal-to-noise ratio and at high bandwidth, which opens the door to a variety of single-molecule biophysics studies.
\end{abstract}

\vspace{2pc}

\section{Introduction}

Zero-mode waveguides (ZMWs) are sub-wavelength apertures in metal films \cite{Zhu2012AnnualReviewofBiophysics}. One of their major properties is that light transmission through the pore is strongly suppressed. This was first described by Bethe for an ideal case of a perfectly conducting sheet \cite{Bethe1944PhysicalReviews}, and subsequently refined by numerical calculations for real materials \cite{Zhu2012AnnualReviewofBiophysics}. A cutoff wavelength \(\lambda_c\) can be defined for ideal ZMWs, being the largest wavelength that is still transmitted. For \(\lambda > \lambda_c\), the impinging light forms an evanescent field inside the pore (figure \ref{fig:1}a) with an exponential intensity decay: \( I(z) = I_0 \mathrm{e}^{-2z \sqrt{\frac{1}{\lambda_c}^2 + \frac{1}{\lambda}^2}} \), where \(I_0\) is the intensity at the surface, \(z\) the height inside the ZMW, and \(\lambda\) the light's wavelength in the medium. The cutoff wavelength can be computed to be \(\lambda_c = 1.7d \), with \(d\) the diameter of the pore \cite{levene2003zero}. Thus for pores much smaller than the wavelength used, this intensity decay can be approximated as \(I(z) = \mathrm{e}^{-2z \sqrt{\frac{1}{1.7d}^2 + \frac{1}{\lambda}^2}} \approx  \mathrm{e}^{- \frac{2}{1.7d} z}\).

Since their first experimental realisation \cite{levene2003zero}, zero-mode waveguides have been applied in several fields such as Fluorescence Correlation Spectroscopy (FCS) \cite{levene2003zero,Samiee2005BiophysicalJournal}, fluorescence enhancement \cite{Rigneault2005PhysRevLett, Gerard2008Phys.Rev.B, Aouani2009ACSNano}, DNA sequencing applications \cite{wanunu2017nnanotech}, and translocation studies \cite{meller2016advmat,Auger2014Phys.Rev.Lett.}. In most applications, ZMWs are built directly on glass, whereas in order to enable translocations, they need to be fabricated within a freestanding membrane. Both gold \cite{Gerard2008Phys.Rev.B, Aouani2009ACSNano, meller2016advmat} and aluminum \cite{levene2003zero, Samiee2005BiophysicalJournal,Rigneault2005PhysRevLett, foquet2008aplphys} are widely used as metals for ZMWs, and a range of different fabrication techniques have been employed, such as lift-off \cite{meller2016advmat,foquet2008aplphys}, dry etching \cite{levene2003zero}, and Focused Ion Beam (FIB) milling \cite{Rigneault2005PhysRevLett,Baibakov2019ACSNano}.
Whereas for FCS applications a decrease of the detection volume is exploited, a main reason to use ZMWs in other applications is that they allow a sharp localization of the detection area which both reduces bleaching and suppresses unwanted background. 

In this work, we aimed to develop a platform that allows to study the translocation behavior of freely diffusing fluorophores, proteins, and DNA, in at least two different color channels. Optical detection of translocating molecules provides benefits compared to ion-conductance-based nanopore detection of DNA and proteins \cite{meller2016advmat} that can be limited due to low signal-to-noise ratio at low salt and low voltages \cite{Fragasso2020ACSNano}. Moreover, bias-free diffusion can not be studied by conductance-based techniques, as the applied voltage  that is a prerequisite for the read-out current signal simultaneously serves as a driving force.

Here, we first describe how freestanding palladium ZMWs can be fabricated and why they exhibit advantages compared to ZMWs made out of other metals. Next, we demonstrate that translocations of freely diffusing fluorophores of two colors can readily be detected. Subsequently, we show that the translocation of proteins can be quantified using our method. Finally, we make clear that not only freely diffusing systems can be investigated, but Pd ZMWs can also be used to detect voltage-driven DNA translocations in regimes that are inaccessible by conductance-based nanopore measurements.

\section{Methods}

\subsection{Fabrication of Freestanding Zero-Mode Waveguides}

Freestanding SiN membranes of \SI{20}{nm} thickness were fabricated  following the protocol from \cite{Janssen2012Nanotechnology}. After a cleaning step of \SI{2}{min} oxygen plasma at \SI{100}{W} at a flow rate of \SI{200}{ml\per min} in a PVA Tepla 300 plasma cleaner, \SI{5}{nm} chromium was deposited onto the SiN layer at \SI{0.5}{\angstrom \per sec} under a base pressure of \SI{3E-6}{torr} in a Temescal FC2000 e-gun evaporator. Immediately thereafter, without breaking the vacuum, a \SI{100}{nm} layer of Pd was evaporated at \SI{1}{nm\per sec} at a base pressure below \SI{2E-6}{torr}. Subsequently, the SiN membrane was removed by dry etching in a Leybold Fluor F2 dry etcher at \SI{50}{W}, \SI{50}{sccm} CHF\textsubscript{3} and \SI{2.5}{sccm} O\textsubscript{2}, which produced \SI{8.6}{\micro \bar} of chamber pressure. The etch rate at these settings was about \SI{2}{\angstrom \per \second}, so the SiN layer was heavily overetched by the \SI{3}{min} etching time. This overetching was done in order to thoroughly remove the SiN. In a next step, the chips were submersed in isopropylalcohol and deionized water in order to remove any bubbles trapped at the membrane, and then submersed in Cr etchant (Microchemicals TechniEtch Cr01) for \SI{30}{s} to remove all the Cr from the window. The etch rate is known to be about \SI{60}{nm \per min}. After a wash in  H\textsubscript{2}O and IPA the samples were spin dried.

Subsequently, the chips with the freestanding Pd films were transferred to a FEI Helios G4 CX Focused ion beam/SEM, where ZMW holes were drilled at different pore sizes using a \SI{2}{pA} Ga beam at a  \SI{30}{kV} acceleration voltage. Stigmation correction of the ion beam turned out to be crucial to produce reproducible results for pore sizes below \SI{40}{nm}. Markers were fabricated by locally thinning down the Pd membrane while not piercing through. The results were immediately checked using the immersion mode of the SEM and later in a JEOL JEM-1400 TEM in order to accurately measure the pore size for each pore used in the study.

\subsection{Imaging of Biomolecules}

Freestanding Pd membranes were mounted in a custom-built flow cell \cite{Keyser2006ReviewofScientificInstruments} that allows for selective flushing of solutions to the reservoir and the detection side. These flow cells also  allow to simultaneously apply a voltage across the membrane, flow solutions along the membrane, and image the membrane with an optical microscope. After a \SI{2}{min} oxygen plasma at \SI{50}{W}, done in order to increase the hydrophilicity, the chips were mounted such that the Pd membrane side faced the microscopes objective. Wetting of the pore was checked by conductance measurement using Ag/AgCl electrodes and an Axopatch 200B amplifier (Molecular Devices). The flowcells were then mounted onto the stage of a Picoquant MicroTime 200 confocal microscope. A flow on the detector side was generated using a syringe pump at \SI{2}{ml/hr} to prevent the  accumulation of fluorophores near the ZMWs on the detection side. The marker positions were determined by the transmitted light and the lasers were focused in the center of the markers crossing (\ref{fig:1}d,f), where the nanopore is located (\ref{fig:1}e). For biomolecular-translocation studies, \SI{640}{nm} and \SI{485}{nm} lasers were used in continuous wave mode at \SI{240}{\micro W} and \SI{55}{\micro W} focus power, respectively. The lasers were focused using a 60x Olympus UPSLAPO 60XW water immersion objective with a working distance of \SI{280}{\micro m}. The emission light was split by a dichroic mirror and was filtered by either a 690/70 or 550/88 optical filters before being detected by a single-photon avalanche-diode detector.

For DNA translocation studies, \SI{150}{mM} KCl solution buffered to pH 8.0 with 1xTE (Sigma) was used on both sides. DNA of two different sizes, 50 basepairs (bp) or 1 kilobasepairs (kbp) (NoLimits ThermoFisher), was added at \SI{5}{ng/\micro \liter} with \SI{100}{nM} of Yoyo1 intercalating dye. Alexa Fluor 647-labeled BSA (Sigma) and Atto dyes (Atto Tec) were diluted in 1xPBS with 1mM MgCl\textsubscript{2}. Analytes were flushed to the reservoir  side (cis) and detected after translocation through the ZMW to the membrane side (trans). For DNA translocation studies, a voltage was applied across the membrane.The exact position of the pore was located and lasers were focused on it, upon which time traces of the fluorescence intensity  were recorded for several minutes for each condition.

Diffusion constants, Atto647N concentrations, and Atto488 concentrations were measured by FCS in open solution in the respective buffer on the same PicoQuant microscope and analyzed using SymphoTime software. Protein concentrations were determined using a DeNovix DS11 Spectrophotometer. DNA was diluted directly from the stock.
In order to reuse the device, the PDMS and acrylic parts of the flowcell were cleaned with DI water and ethanol  after each experiment. The chips were cleaned with ethanol, IPA, and acetone, and the O-rings and PEEK parts of the flow cell were cleaned  in 30\% H\textsubscript{2}O\textsubscript{2} in order to remove any remaining fluorophores.

\subsection{Event Detection and Data Analysis}

Photon-arrival time traces were analysed with a custom-written python script using several packages \cite{Hunter2007ComputinginScience, Virtanen2020NatureMethods, McKinney2010, vanderWalt2011ComputinginScienceEngineering, Perez2007ComputinginScienceEngineering}. The code is freely available at \cite{codepdzmwpaper}. All the data used for this study were deposited in the open photon-hdf5 format \cite{Ingargiola2016BiophysicalJournal} and can be downloaded from \cite{datapdzmwpaper}.

Photon bursts, indicating the times when a single fluorophore was present in the laser focus, were detected using an algorithm developed by Watkins and Yang \cite{Watkins2005TheJournalofPhysicalChemistryB} that is based on detecting change points in the photon-arrival statistics by applying generalized-likelihood-ratio tests on the time trace. The software can be found at \cite{hawyang2020}. The advantage of using the time-resolved photon data for burst detection lies mainly in the avoidance of any \textit{a priori} binning of the data. This is beneficial because artifacts can be introduced by applying binning when signals vary in time scales. We set the false positive level to \(\alpha = 0.01 \) which means that 1 out of 100 photon bursts detected by the algorithm will be a false positive. For long time traces, we encountered a substantial amount of these, and therefore in a next step, we discarded bursts that were below a certain threshold in the amount of photons, as shown in supplementary figure \ref{fig:s_threshold}. This threshold function was empirically determined from  time traces where no fluorophores were present in the system. We found that false positive bursts  typically had a very long duration and a photon rate that was only slightly above background level. Additionally bursts that involved less than 4 photons were discarded.

\section{Results}

First in section \ref{sec:pdzmw}, we describe how Pd ZMW devices were fabricated and used for translocation experiments. In section \ref{sec:free_fluo}, we demonstrate two-color detection of individual fluorophores, that freely diffuse through the ZMW. In order to show the feasibility to work with biomolecules, we subsequently  investigate the translocation behavior of proteins in section \ref{sec:proteins}. After that, we show that even voltage-driven translocations are possible with Pd ZMWs by demonstrating DNA translocations in section \ref{sec:dna}.

\subsection{Palladium is a Suitable Metal for Freestanding Zero-Mode Waveguides}
\label{sec:pdzmw}

We identified that palladium is an excellent metal for the fabrication of freestanding ZMWs, due to its chemical stability, low optical background, and mechanical stability. We started out to develop a system to study biomolecular translocations through a ZMW nanopore  with many requirements: Most prominently,  it should be possible to detect translocations of both DNA and protein at a high signal-to-noise ratio. Next, the detection efficiency should be irrespective of a driving force and buffer conditions. Additionally, different types of proteins should be distinguished when labeled differently. When working with proteins, the possibility to work with physiologically relevant buffers such as PBS is crucial. When working with DNA it is beneficial to apply a voltage across the pore. Finally, we chose to use an optical approach and started out to fabricate freestanding ZMW structures (see Methods) in order to develop such a system.

In the past,  gold \cite{Gerard2008Phys.Rev.B, Aouani2009ACSNano, meller2016advmat} and aluminum \cite{levene2003zero, Samiee2005BiophysicalJournal,Rigneault2005PhysRevLett, foquet2008aplphys} were typically used as substrate metals for ZMWs. Whereas Au is chemically more stable, Al has better optical properties \cite{Zhu2012AnnualReviewofBiophysics}. More specifically, base metals such as aluminum are not very stable in chlorine-based solutions that are commonly used for nanopore ion conductance measurements. Hence for long-lasting translocation experiments, aluminum ZMWs need to be chemically protected, e.g.\ by applying a SiO\textsubscript{2} layer \cite{wanunu2017nnanotech, MoranMirabal2007Nanotechnology, Larkin2014NanoLetters}, but this can be a difficult process, and disadvantageous when the interior surface of the ZMWs needs to be coated.

\begin{figure}[h]
    \centering
	\includegraphics[width=0.45\textwidth]{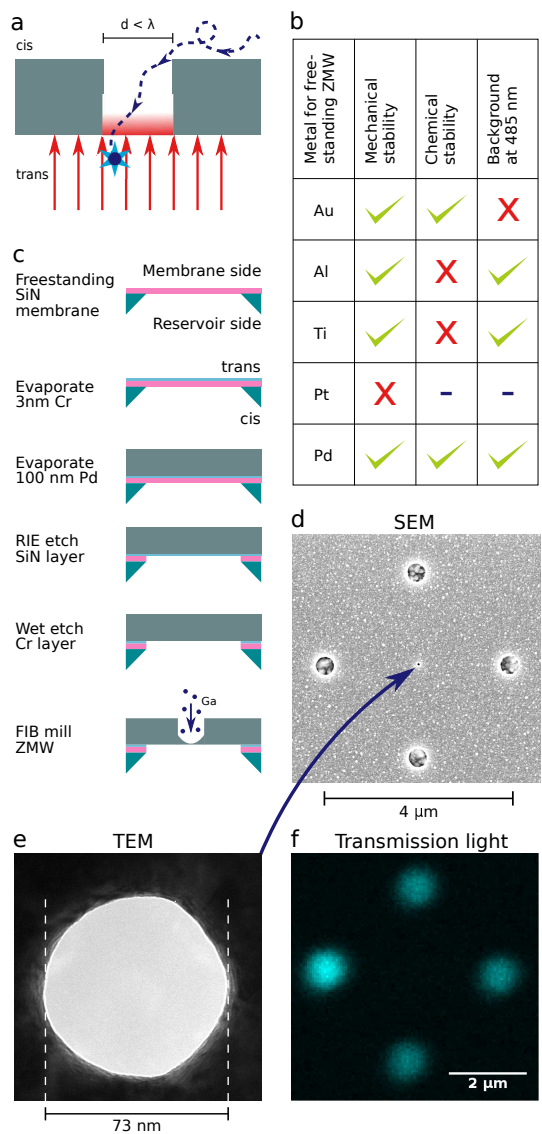}
	\caption{\textbf{Fabrication of Pd ZMW} 
	a: A zero-mode waveguide is a sub-wavelength aperture in a metal film. They have the property to block impinging light such that an evanescent field is generated in the pore with an exponential intensity decay. When a fluorophore traverses the nanohole from the cis side towards the laser-illuminated trans side (dashed line), it can be detected by its fluorescence. b: Various metals were tested for their suitability for ZMW detection of translocating single molecules. Whereas gold produced a background in the \SI{550}{nm} range when illuminated with a \SI{485}{nm} laser, aluminum and titanium were not stable in chlorine-based solutions, see SI Figure \ref{fig:s_tipore}. When platinum was deposited, its mechanical stress made it impossible to fabricate free-standing membranes. Only palladium fulfilled all requirements. c: Freestanding Pd films were made by e-beam evaporation and support layers were subsequently removed by etching. Local ZMW and marker structures were fabricated using Focused Ion Beam Milling. d: SEM image of the marker structures, that allow a localization of the nanohole. e: TEM image of the central nanopore in d. f: Transmission light image of the markers. \label{fig:1}}
\end{figure}

We tested Au, Al, Ti, Pt, and Pd for their performance as materials suitable for ZMWs, see figure \ref{fig:1}b. Whereas in principle Au was good to work with, we encountered a high optical background signal in the \SI{550}{nm} channel, when the gold surface was illuminated with a \SI{485}{nm} laser. This background intensity was present when the laser was focused either on the membrane or on the ZMW. The optical background intensity was similarly detected when focusing on a single crystalline gold surface. Unfortunately, the region around \SI{550}{nm} is in the detection band for common fluorophores such as Atto488 and Alexa 488, hence rendering gold less suitable for our application. Whereas using fluorophores of lower wavelengths would even increase the background, fluorophores at longer wavelengths start to overlap with the \SI{640}{nm} channel. Since we aim for well-separated multicolor detection, we conclude that gold was a less suitable choice for the metal.

Whereas Al and Ti both had desirable optical properties,  pores of these materials  were found to be instable during the time of experiments as they changed their shape or closed. For Ti this was especially prominent when a voltage was applied. Examples are given in supplementary figure \ref{fig:s_tipore}. Passivation with e.g.\ SiO\textsubscript{2} might in principle provide a solution but can present a range of difficulties, and hence we tried more noble metals such as platinum and palladium. When we evaporated Pt onto our SiN substrates, the resulting films were found to be under such a high stress that the membranes broke  quickly during handling, so we could not investigate its properties further. Our last candidate Pd, however, fulfilled all the requirements, as that we could fabricate  freestanding ZMWs that were chemically and mechanically stable and had  a low optical background, as described in figure \ref{fig:1}c. 

For fabricating these Pd ZMWs, a Pd film was deposited on a freestanding SiN membrane on a Si support, and the underlying SiN was subsequently removed, leaving only a freestanding Pd membrane of \SI{100}{nm} thickness.  We chose for Focused Ion beam (FIB) milling in order to drill the ZMW structures. The big advantage of FIB compared to lithography-based techniques is that it is straight-forward to fabricate structures of different heights within the same metal film. This property was used to make shallow markers around a central ZMW that was a small throughhole (see figure \ref{fig:1}d), which enabled us to locate the exact position of the nanohole. This was necessary as the ZMW of less than \SI{100}{nm} diameter is not visible in the optical microscope, neither in scattered light, nor in transmission light. An example of an optical image of the marker pattern is given in figure \ref{fig:1}f where the structure is shown in transmission light.

With this method we were able to fabricate ZMWs in freestanding Pd films of \SI{100}{nm} thickness, that were both chemically and mechanically stable enough to serve for experiments in buffers such as PBS for several hours under illumination of a \SI{640}{nm} and \SI{485}{nm} laser, while providing only a low optical background.

\FloatBarrier
\subsection{Two-Color Detection of Freely Diffusing Fluorophores}

We aimed to detect single fluorophores that were freely diffusing through these Pd ZMWs. The  bottleneck in such measurements is the amount of photons that are detected from a single-molecular analyte. Freely diffusing fluorophores pose the biggest challenge to the system, since every other analyte of interest would spend more time in the laser focus because of a lower diffusion constant due to a higher mass. Additionally, they might be labeled with more than one fluorophore per molecule. The diffusion constant highly influences the photon yield as the main detection happens in the laser focus volume after the translocation of the analyte through the ZMW nanohole. A higher diffusivity  leads to a shorter dwell time and hence less photons emitted from the analyte when it diffuses out of the laser focus volume.  Note that the detection region in our approach is different from typical applications of ZMWs on glass slides, where an enhancement in FCS signal is due to a volume reduction such as in \cite{levene2003zero}.

\begin{figure}[h]
    \centering
	\includegraphics[width=\textwidth]{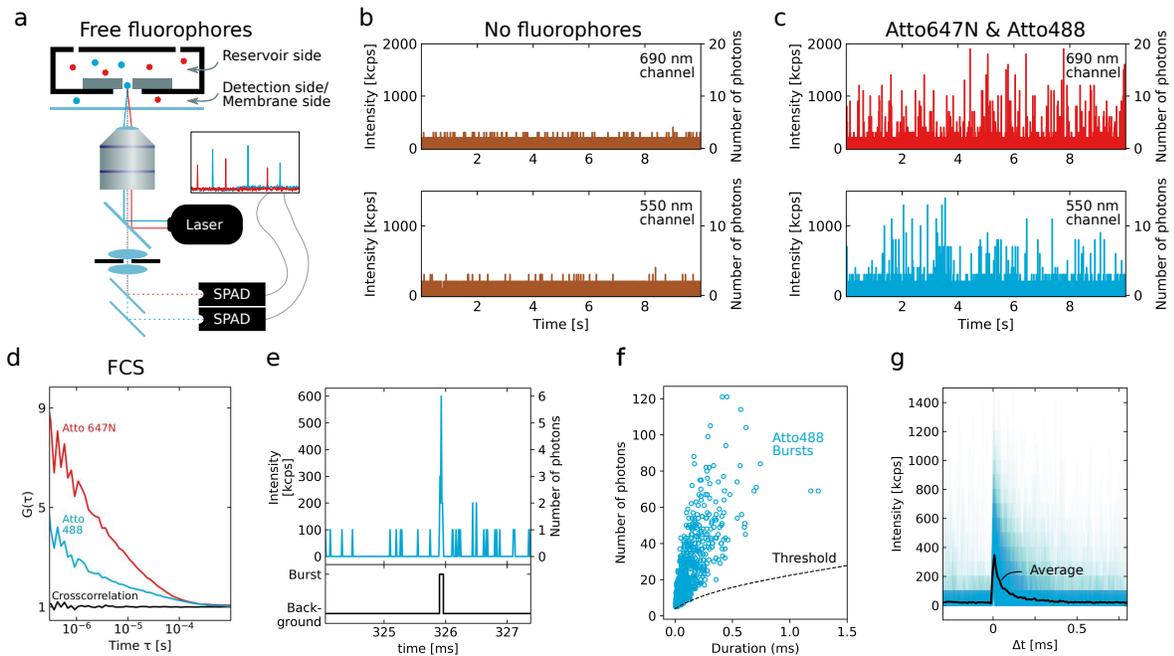}
	\caption{\textbf{Free fluorophore translocations} Freely diffusing fluorophores are well resolved, despite the fact that they pose the biggest challenge to the system, due to their high diffusion constant and their low brightness.  a: Schematic of the setup. A Pd ZMW device is mounted in a flowcell, that enables the specific flushing of liquid to the reservoir side (top) or to the membrane side (bottom) of the device\cite{Keyser2006ReviewofScientificInstruments}. The flowcell is mounted on a confocal microscope where multiple lasers can be focused on the nanopore. In the detection pathway, the two spectral channels are split, filtered, and detected by single-photon avalanche diode detectors that record the arrival times of individual photons. b: Time traces of the optical background when no fluorophores are present. The background rates amount to about \SI{10}{kcps} and \SI{8.5}{kcps} in the \SI{485}{nm} and \SI{640}{nm} channel, respectively. c: Time traces when \SI{7}{nM} of Atto488 and \SI{4}{nM} of Atto647N are added to the reservoir side. Translocations are clearly visible. 
    d: Autocorrelation and crosscorrelation curves for the two spectral channels. The low crosscorrelation shows that the spectral separation works adequately. e: Illustration of the spike-detection algorithm \cite{Watkins2005TheJournalofPhysicalChemistryB}: Individual photon bursts (bottom) are found from the raw data (top), irrespective of data binning. When lots of photons appear within a short time frame, a burst (due to a fluorophore entering the laser focus) is identified. f: The amount of photons in a burst vs.\ its duration. Bursts with less photons than the threshold were removed.  g: Typical burst shape (black) from averaging of 777 individual bursts (blue). \(\Delta t\) is the time to the onset of the burst. Binning in b, c, e, and f is \SI{10}{\micro s}. \label{fig:2}}
\end{figure}

In order to detect free fluorophore translocations, the chips containing the freestanding Pd membrane were mounted in a flowcell that allowed selectively flushing the reservoir (cis) and the membrane side (trans) with solutions \cite{Keyser2006ReviewofScientificInstruments}.  For this first study, we chose to use a confocal microscope with multiple lasers. Fluorescent light was collected with two single-photon avalanche diode detectors, which recorded the photon traces in two separate spectral channels. These can detect a photon every \SI{50}{ns} and thus are not limiting the time resolution. As they record the arrival time of each photon, they allow for more elaborate spike-detection algorithms than  common binning and thresholding methods.  The laser power in the focus used throughout all  experiments was determined by optimizing the signal-to-noise ratio in each channel to be  \SI{240}{\micro W} of the \SI{640}{nm} laser  and \SI{55}{\micro W} of the \SI{485}{nm}, with both lasers driven in continuous wave mode. A bare Pd ZMW without any fluorophores in the system (figure \ref{fig:2}b), yielded an average count rate of \SI{8.5}{kcps} in the \SI{690}{nm} channel and \SI{10}{kcps} in the \SI{550}{nm} channel. This means that on average \(\sim\)10 photons per millisecond were recorded in the backscattered light from the Pd. In other words, in order to reliably detect a fluorophore, its count rate needs to be significantly higher than this background value.

We were able to clearly resolve single fluorophores of Atto647N and Atto488, as shown in the time traces in figure \ref{fig:2}c. Furthermore, both channels could be mutually well discriminated: Whereas each channel showed a great degree of autocorrelation, the crosscorrelation between the channels was flat (figure \ref{fig:2}c), indicating very little crosstalk between the two channels and a good spectral separation. The shape of the FCS curves differed from the one expected for freely diffusing fluorophores by showing a less step-like decrease in autocorrelation over time. We attribute this mainly to the nearby surface.

As we had access to the single photon times, we were able to perform a burst-detection analysis based on the photon arrival statistics, using the algorithm developed by Watkins and Yang \cite{Watkins2005TheJournalofPhysicalChemistryB}. In brief, this algorithm detects statistically significant changes in the arrival times of photons and thus  sorts a time trace into different states. Here, one state was the background state with a low photon rate and another state was the event state where many photons arrive within a short time (Fig.2e). Typically, in a minute-long time trace that records the background signal, \(\sim\)600,000 photons are involved. Due to this large number, even at a false positive rate of 1\% a certain amount of photon bursts are identified which are actually not associated with a fluorophore in the detection area. This can disturb translocation rates, especially for time traces where few fluorophores are present in the system. Therefore we applied an empirical threshold function to remove most of these false positives (see Methods). In brief, it discards events with less than 4 photons and compares the photon rate for long events to the background rate obtained from measurements where no fluorophores were present. The threshold function is shown as a dashed line in figure \ref{fig:2}f.

When looking closer at the average shape of the fluorescence bursts (figure \ref{fig:2}g), several properties can be identified. There is a quick onset of fluorescence in the beginning of the burst to an average photon rate of about \SI{400}{kcps}, which can be explained as the fluorophore exiting from the ZMW into the laser focus. Then, the intensity  decays within less than \SI{100}{\micro s}, which is due to the fluorophore leaving the laser focus or bleaching. After \SI{500}{\micro s} after the burst, no trace of additional fluorescence can be seen in the average photon rate. This time gives an upper bound for the dwell time of the fluorophore in the laser focus. Concretely, if the time interval between two bursts exceeds \SI{500}{\micro s}, they likely arise from two different fluorophore translocations. Accordingly, the maximum detectable translocation event rate is about \SI{2}{kHz}.

\label{sec:free_fluo}
\FloatBarrier

In this section we showed that using Pd as a substrate for freestanding ZMW, the translocations of freely diffusing fluorophores of two different wavelengths can simultaneously be detected. A major factor in the success of these experiments was the choice of Pd as a substrate material, as it is mechanically and chemically stable enough for  several hour-long  experiments. Additionally it has a low optical background when illuminated with a blue laser.

\subsection{Protein translocations can be resolved and quantified}
\label{sec:proteins}

Characterization of single proteins is of great interest \cite{Ledden2011}. Protein translocation has  been attempted before in nanopore measurements of the ionic conductance, with rather modest success however \cite{plesa2013nanol}. A major difficulty in ionic-conductance probing of protein translocations is the high speed at which proteins  traverse solid-state nanopores, which makes that almost all proteins escape detection.  
Here, we benefit from the outstanding temporal resolution of the optical detection to resolve virtually all proteins that traverse the nanopore. We  measure translocations of  Bovine Serum Albumin (BSA) labeled with Alexa647, as it is a standard protein that is  widely used in biological research.

\begin{figure}[h]
    \centering
	\includegraphics[width=\textwidth]{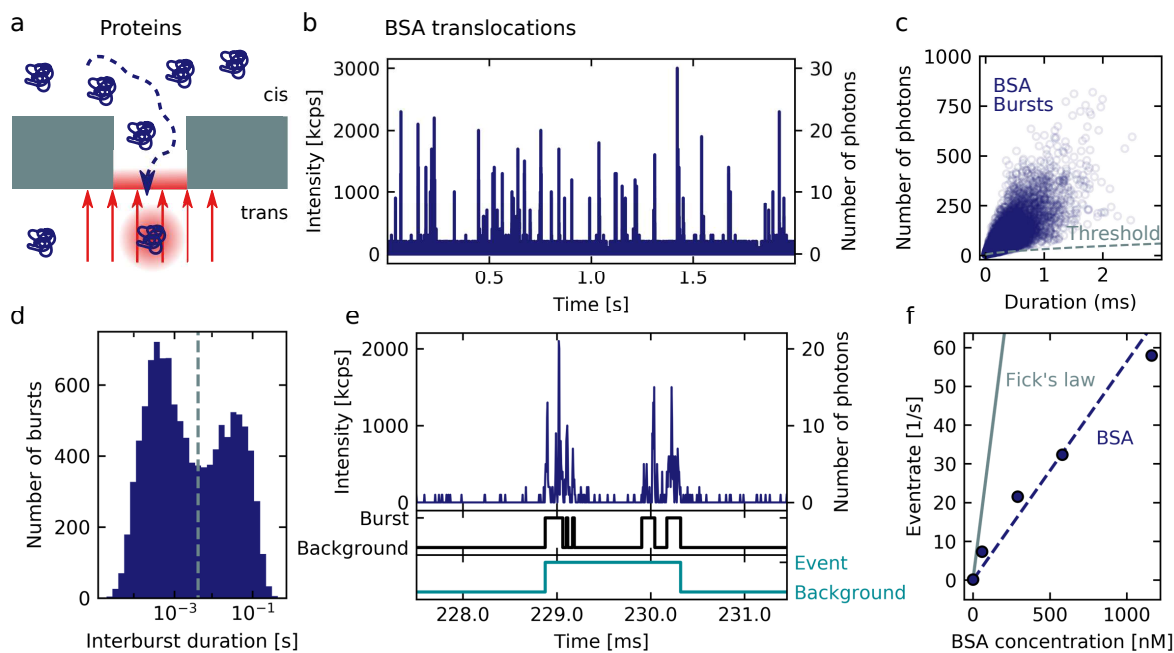}
	\caption{\textbf{Diffusion Driven Protein Translocations} a: Schematic of protein translocations. The translocation of fluorophore-labeled proteins can be studied using Pd ZMWs, without requirements for a specific salt concentration or a bias voltage. b: Time trace of \SI{290}{nM} \textit{bovine serum albumine} (BSA) labeled with Alexa647 that freely translocate through a \SI{43}{nm} pore. c: Number of photons per burst vs.\ burst duration. Bursts below the threshold likely do not involve a fluorophore-labelled protein and were removed. The full scatter plot is shown \ref{fig:s_add}a.  d: Number of bursts versus the interburst duration. Two populations can be identified. The populations with the lower interburst duration is attributed to re-entries of the same protein to the laser focus. e: One example event that illustrates how multiple bursts are combined to individual long events when their interburst duration is below a certain threshold time (here \SI{4}{ms}). f: Translocation rate versus BSA concentration. Error bars (standard deviation due to the statistical error) are smaller than the data markers. Solid grey line shows the prediction by Fick's law of diffusion. The dashed line is a linear fit to the data. Binning in b and e is \SI{10}{\micro s}.  \label{fig:3}}
\end{figure}

As shown in figure \ref{fig:3}b, a continuous trace of fluorescence bursts was detected upon addition of \SI{290}{nM} of BSA to the reservoir cis side. We can compare the burst duration with an estimate from pure diffusion based on \( \langle x^2 \rangle = 2Dt \). Using BSA's diffusion constant of \SI{52}{\micro\metre^2\per\second} and a diffraction limited laser focus  radius of \SI{200}{nm} gives about \SI{0.4}{ms}, which compares well with the measured dwell times \ref{fig:3}c. In order to assess the translocation rate, we first have a closer look at the succession of individual fluorescence bursts. The interburst duration is defined as the time between the end of a fluorescence burst and the beginning of the next. When plotting this in a histogram, two populations can be identified, see figure \ref{fig:3}d. This distribution shows a minimum at about \SI{4}{ms}. We interpret the population with the lower interburst duration as re-entries of the same protein to the laser focus. This also becomes clear when looking at the example time trace in figure \ref{fig:3}e. Whereas the burst-detection algorithm detects 5 bursts separated by short gaps, it is very likely that the entire trace actually results from one translocating protein only. Therefore in a second step of the detection algorithm, we assigned all bursts with an interburst duration of less than \SI{4}{ms} to single translocation events. The actual threshold for the combination (here 4 ms) depends  on the diffusion constant and has to be determined from low-concentration experiments for each analyte individually. This number also determines the maximum rate at which we can distinguish the succession of individual protein translocations, i.e. the maximum event rate that we can resolve, which for BSA would be \SI{250}{Hz}. 

Armed with this analysis, we calculated the translocation rate for different BSA concentrations (figure \ref{fig:3}f). The rate shows an approximately linear dependence on concentration, as is expected from diffusion, based on Fick's law \cite{Muthukumar2014TheJournalofChemicalPhysics}. The measured translocation rates are lower that those calculated  based on the pore size and the protein concentration, by  a factor of 4.6. This  is more than 2 orders of magnitude better than what is common for conductance-based nanopore experiments for BSA translocations  where typically a very large quantitative discrepancy is observed \cite{plesa2013nanol}. In the calculation of the expected translocation rate a reduction of the diffusion constant due to the spatial confinement in the pore, as reported by Dechadilok and Dean \cite{Dechadilok2006Industrial&EngineeringChemistryResearch}, was included. We did however not include any corrections for surface protein interaction as suggested by Muthukumar \cite{Muthukumar2014TheJournalofChemicalPhysics} as their strength is  hard to estimate. Other reasons for the remaining small discrepancy might be a loss of protein in the upper part of the flowcell.

\FloatBarrier

\subsection{Voltage-Driven DNA Translocations}
\label{sec:dna}

\begin{figure}[h]
    \centering
	\includegraphics[width=\textwidth]{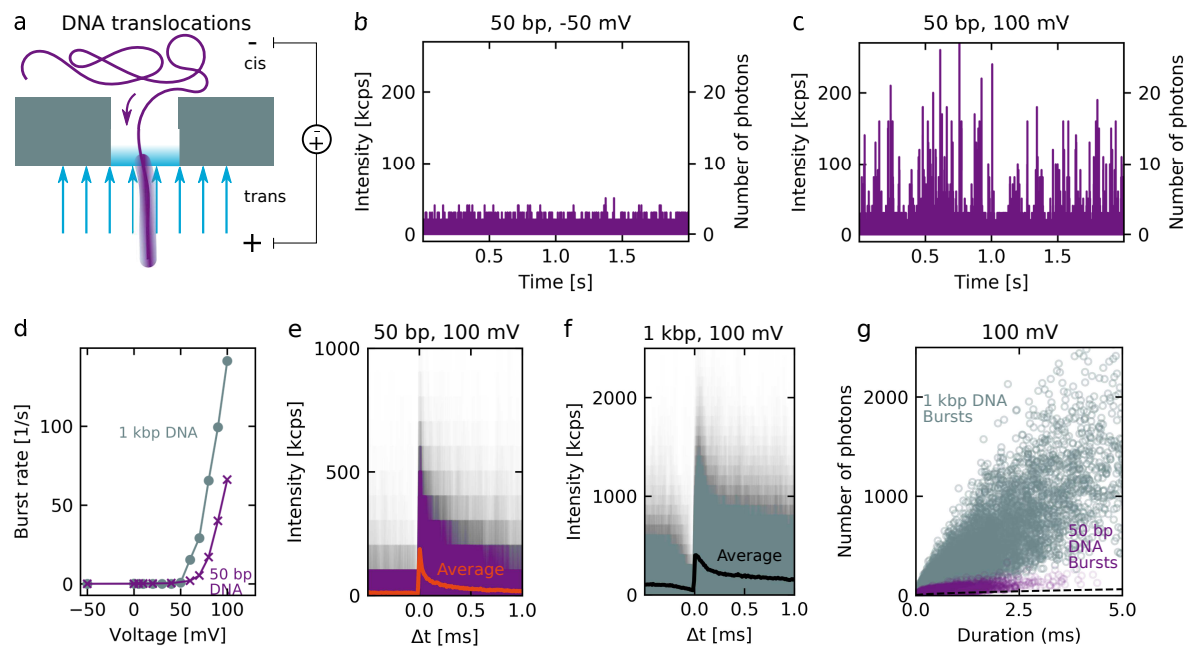}
	\caption{\textbf{Voltage Driven DNA Translocations} a: DNA can be driven  through a ZMW by a voltage that is applied across the membrane, where  individual translocation events are recorded optically. b: When a negative bias of \SI{-50}{mV}is applied to the trans side, no translocations  of \SI{50}{bp} double-stranded DNA labeled with the intercalating dye Yoyo1 were observed through the \SI{59}{nm} by \SI{47}{nm} oval pore. c: At \SI{+100}{mV}, however, many events were detected. Notably these DNA translocations are well resolved despite the low salt, low bias voltage, short DNA length, and large pore size. d: Burst rate versus driving voltage for two different DNA lengths. Errorbars  (standard deviation due to the statistical error)  are smaller than the markers. e: Average shape from 2000 individual fluorescence bursts for 50 bp DNA. f: Same for 1 kbp DNA. g: Photons per burst vs.\ burst duration at \SI{+100}{mV} bias. Bursts below the dashed line are ignored in further analysis. Bursts from longer DNA involved more photons. The full dataset is displayed in supplementary figure \ref{fig:s_add}. DNA concentrations were  \SI{5}{ng \per \micro \litre} in the reservoir cis side. Binning for the plots in b and c was \SI{100}{\micro s} and \SI{10}{\micro s} for e and f. \label{fig:4}}
\end{figure}

After having showed that these Pd ZMWs are useful for detecting analytes which freely diffuse through the pore, in a last step we now show that they can also detect biomolecules which are translocating by being driven through the pore. Specifically, we studied voltage-driven DNA translocations. A voltage was applied across the membrane using Ag/AgCl electrodes that were connected to a patch clamp amplifier and immersed in \SI{150}{mM} KCl solution (pH 8). DNA at \SI{5}{ng \per \micro l} together with \SI{100}{nM} Yoyo1 intercalating dye was added to the reservoir cis side of the chip. A negative voltage of \SI{-50}{mV} was applied to the trans side, which prevented the DNA from translocating. Subsequently, the trans voltage was increased up to \SI{+100}{mV}, yielding  more than 50 fluorescence bursts per second  (figure \ref{fig:4}b,c) in the \SI{550}{nm} channel.

We find  a higher burst rate for DNA molecules of \SI{1}{kbp} length compared to the \SI{50}{bp} fragments, see figure \ref{fig:4}d. Similarly, a higher translocation rate for longer DNA was observed by Wanunu \textit{et al.} in \SI{5}{nm} SiN nanopores \cite{Wanunu2009NatureNanotechnology}. More differences between the detection of the short and the long DNA are visible when looking at the average burst shape. In figure \ref{fig:4}e and \ref{fig:4}f, 2000 individual fluorescence bursts are overlaid and their average fluorescence intensity is displayed. These show, together with the scatter plot in figure \ref{fig:4}g, that the short DNA has, as expected, a shorter dwell time in the laser focus as well as a lower peak intensity. The shorter dwell time is readily explained by the higher diffusion constant of shorter DNA. Additionally, the \SI{1}{kbp} DNA carries more fluorophores, as the amount of fluorophores that the DNA incorporates scales with the DNA length, which explains the higher peak intensity. This shape of bursts, with a strong onset at the start of the event and  a subsequent gradual  decrease of fluorescence intensity, is similar to the one described  by Assad \textit{et al.} \cite{meller2016advmat}.

Summing up, we could show that freestanding Pd ZMWs are an excellent   system to study DNA translocations, which was not possible for e.g.\ Ti-based systems due to instabilities upon application of a voltage. Notably, DNA translocations could be well resolved  for DNA strands as short as \SI{50}{bp} at low salt, at low driving voltage, and for a very large pore size. All of these are conditions that are very challenging for conductance-based nanopore readout on solid-state nanopores.

\FloatBarrier

\section{Conclusions and Outlook}

Virtually all studies involving zero-mode waveguides used either gold \cite{Gerard2008Phys.Rev.B, Aouani2009ACSNano, meller2016advmat} or aluminum \cite{levene2003zero, Samiee2005BiophysicalJournal, Rigneault2005PhysRevLett, foquet2008aplphys} as metals to block the incident light. It is well established that gold offers more chemical stability and biocompatibility, whereas aluminum has better optical properties  \cite{Zhu2012AnnualReviewofBiophysics}. In order to circumvent the limitations of these metals, we investigated the properties of different metals and identified Pd as the most promising candidate for single-molecule studies. Whereas palladium is well known as a catalyst \cite{Love2005ChemicalReviews}, we only know of studies where Pd was patterned with sub-wavelength apertures for hydrogen sensing \cite{Maeda2012JournalofAppliedPhysics}. Its suitability for single molecule fluorescence experiments in biologically relevant conditions was so far not shown.

With this study we show that Palladium is a material that combines the best of both: Like gold, it is a noble and biocompatible \cite{Love2005ChemicalReviews} metal that is  chemically much more inert than Al. Furthermore its low optical background allows for  detection of individual fluorophores, as we showed in section \ref{sec:free_fluo}. Pd is also an attractive candidate for nanofabrication due to the fact that its grain size is a factor 2-3 smaller than that of gold. In this paper, we presented an easy protocol to fabricate freestanding Pd ZMWs, based on SiN membranes and a method to locate the pores, which are invisible in an optical microscope due to their sub-\SI{100}{nm} size. 

The clear fluorescence bursts that we observed from free fluorophore translocations in section \ref{sec:free_fluo} are comparable to the ones found by Assad \textit{et al.} \cite{meller2016advmat}, for a Cy5 dye in measurements on gold ZMWs with red light.  In our  experiments, we showed that a comparable signal quality can also be achieved in the blue band by using palladium. 

For diffusion-driven translocations of BSA protein through the nanoaperture, we showed that it is possible to detect  the translocation of individual proteins, up to event rates of \SI{250}{Hz}. This was done in a physiological buffer, without the need of a voltage that might alter the translocation behavior. Comparing the observed translocation rates with the prediction due to Fick's law we found a good qualitative consistency and a small quantitative discrepancy of a factor 4.6, which is orders of magnitude better than from previous studies where the translocation rate was measured by use of  ion-conductance blockades through nanopores \cite{plesa2013nanol}. 

In a third step, we showed that DNA translocations could be quantified in voltage-driven experiments with Pd ZMWs. In this study we used low salt (\SI{150}{mM} KCl), short DNA (\SI{50}{bp}), low driving voltages (\SIrange{0}{100}{mV}), and a large-diameter pore (around \SI{55}{nm}). Despite these conditions, which are challenging for the conventional ionic-current nanopore experiments,  we could clearly distinguish the DNA signals from the background light. As we showed that it is possible to detect even single freely diffusing fluorophores, it should not pose any problem  to even resolve single fluorophore-labelled basepairs within the DNA in future experiments. Importantly, our approach separates the detection method from the driving force, which  is a clear advantage of the optical method compared to conductance-based techniques.

The translocation behavior of fluorophores and DNA was studied before using freestanding ZMWs \cite{meller2016advmat,Auger2014Phys.Rev.Lett.,montel2018zmwdna}. To our knowledge, however, no one so far showed  simultaneous multicolor detection of fluorophores, proteins, and DNA, with the additional possibility to apply a voltage across the membrane and thus drive analytes into the detection volume. Indeed, all three groups of analytes could be detected with an excellent signal-to-noise ratio, well above the noise level in our system, which shows the wide applicability of the platform for different research questions, such as studying selectivity of biomimetic Nuclear Pore Complexes  \cite{kowalczyk2011trendsbt}, polymer physics of DNA translocations\cite{Auger2014Phys.Rev.Lett.}, or DNA data storage \cite{Chen2019NanoLetters}.

To extend the capabilities even further, this system could be combined with a nanometer-sized SiN nanopore as was shown before for Au- or Al-based platforms \cite{wanunu2017nnanotech, meller2016advmat}. Additionally to voltage-driven translocations,  the effect of pressure differences on the translocation behavior could also be studied, similar to Auger \textit{et al.} \cite{Auger2014Phys.Rev.Lett.}. Finally, we note that, as an optical technique, it would be straightforward to acquire data from many pores in parallel, allowing for even better statistics.

Summing up, with this work we introduce a new platform to study the translocation behaviour of single biomolecules through nanopores,that  overcomes several limitations of previous ZMWs from other metals and of ion-conductance-based techniques. We foresee many future applications of our approach. 

\section{Acknowledgments}

We thank Jérémie Capoulade for his extensive support with the optical imaging. Additionally we thank Xin Shi, Wayne Yang, Sergii Pud, and Daniel Verschueren for help with  nanofabrication, Sonja Schmid, Biswajit Pradhan, and Alessio Fragasso for discussions, and Tanja Kuhm for help on the manuscript. We acknowledge funding support from the ERC Advanced Grant LoopingDNA (no. 883684) and the NanoFront and BaSyC programs of NWO-OCW. The authors declare no conflict of interest.

\section{References}
\bibliography{thisbib}

\newpage
\appendix
\section{Appendix}

\subsection{False Positive Events are Sorted Out}

\begin{figure}[h]
	\includegraphics[width=\textwidth]{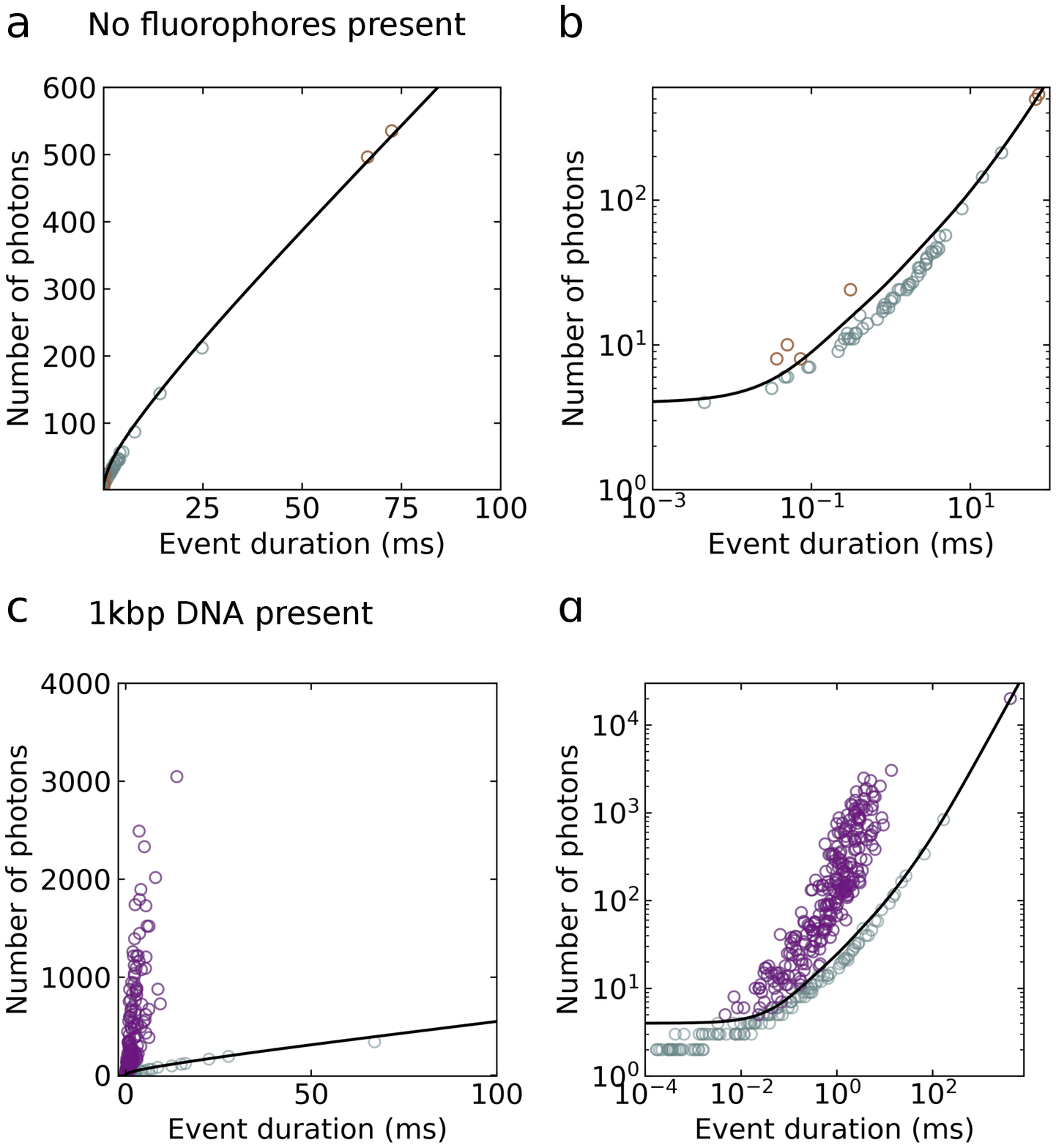}
	\caption{a,b: Scatter plots of the bursts detected by the spike-detection algorithm from the dataset in figure \ref{fig:2}b. If a burst involves less spikes than the empirical threshold  (black line) it is sorted out (grey). In the case of panel a/b, no fluorophores were present in the system, so all remaining bursts (brown) are likely false positives. c,d: Scatter plots from 1 kbp DNA translocation data at \SI{60}{mV}, from figure \ref{fig:4}. The scatter plots show 2 populations, the DNA bursts (purple) and the background bursts (grey) that lie below the threshold (black line). \label{fig:s_threshold}}
\end{figure}
\FloatBarrier

\subsection{Titanium Pores Are not Stable in KCl}

\begin{figure}[h]
	\includegraphics[width=\textwidth]{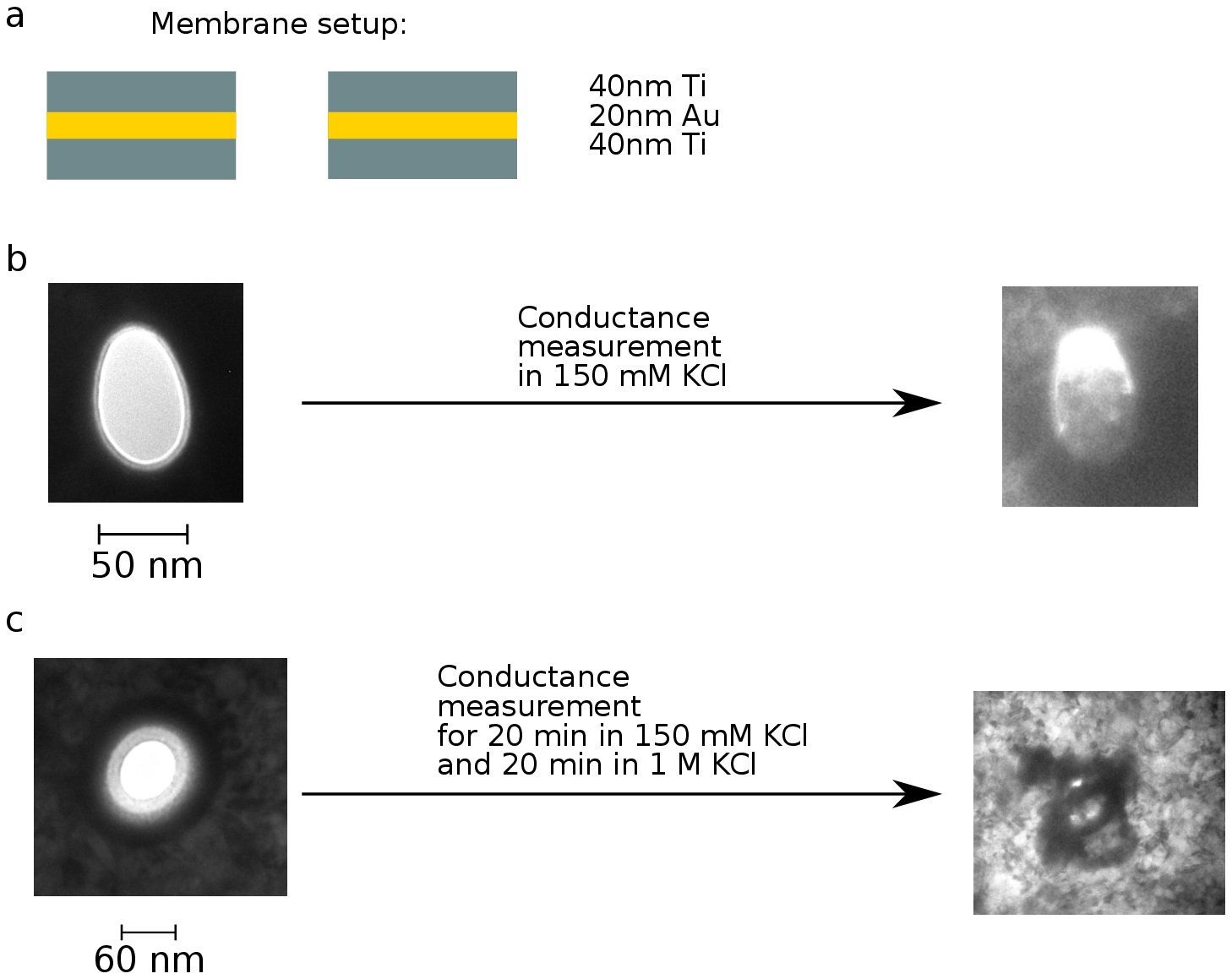}
	\caption{a: During the search for suitable materials we used a Ti, Au, Ti sandwich structure. After fabrication pores had a clear shape (b,c, left side) but after conductance measurements in KCl the pore's structure changed completely (b,c, right side). We concluded that these Ti-Au-Ti sandwich structures were not usable for experiments in liquid. \label{fig:s_tipore}}
\end{figure}
\FloatBarrier

\newpage

\newpage
\subsection{Additional Plots to Figures \ref{fig:3} and  \ref{fig:4}}
\begin{figure}[h]
	\includegraphics[width=\textwidth]{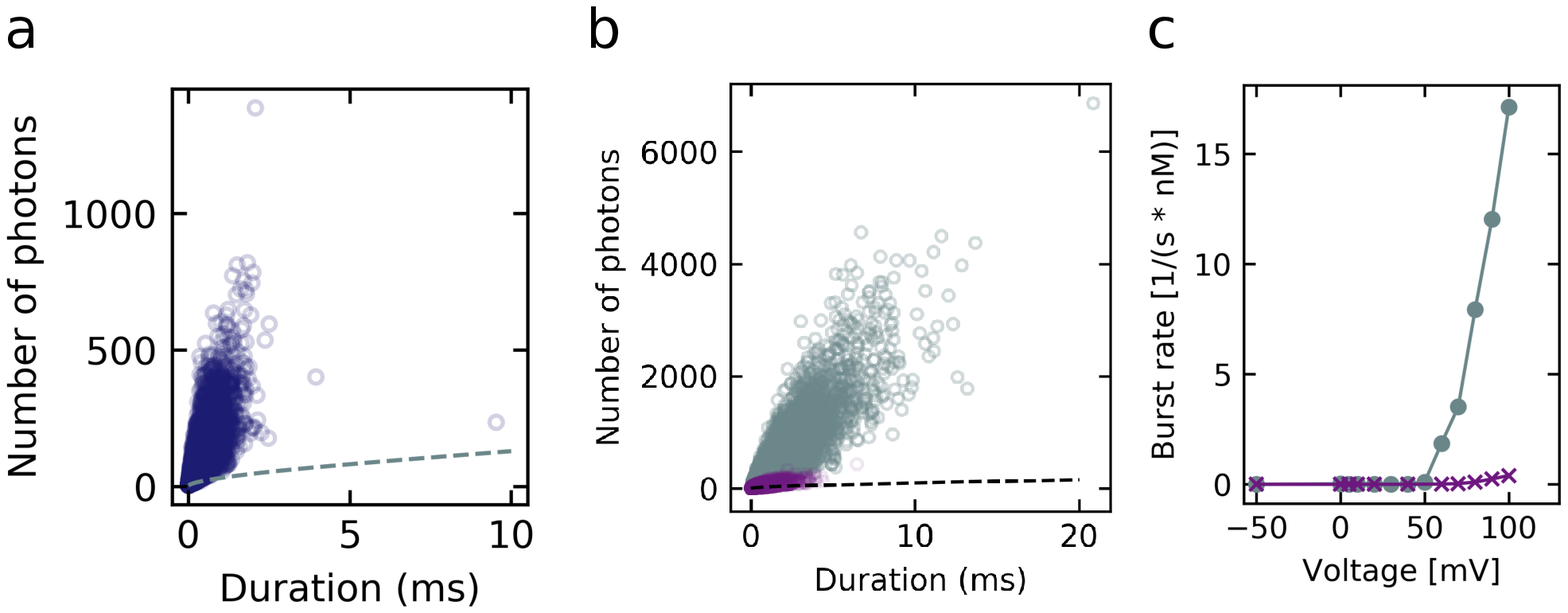}
	\caption{a: Scatter plot of the full dataset shown in figure \ref{fig:3}c. b: Scatter plot of the full dataset shown in \ref{fig:4}g. c: Burst rates shown in figure \ref{fig:4}d, normalized to molarity. \label{fig:s_add}}
\end{figure}

\end{document}

%% file: header.tex
\usepackage[british]{babel}

\usepackage{lmodern}

\usepackage[super]{nth}              %
\usepackage[utf8]{inputenc} %
\usepackage{csquotes}             %
\usepackage{lipsum}			  %

\usepackage{imakeidx}    %
\usepackage [usenames,dvipsnames,svgnames,table] {xcolor} %
\usepackage{transparent}
\usepackage{graphicx}       %
\usepackage{placeins}		%
\usepackage{wrapfig}		%
\usepackage{verbatimbox}	%
\usepackage{tabu}		%
\usepackage{dsfont}         %
\usepackage{siunitx}        %

\usepackage                             {booktabs} %

\usepackage{hyperref}        %
\usepackage{xparse}			 %
\usepackage{xstring}

\DeclareSIUnit\Molar{\textsc{M}}                        %
\hyphenchar\font=\string"7F    %